\documentclass{emulateapj}

\usepackage{psfig}
\usepackage{amsmath}
\usepackage{amssymb}
\usepackage{graphicx}
\usepackage{appendix}

\newcommand{\lSect}[1]{{\label{sec:#1}}}
\newcommand{\lFig}[1]{{\label{fig:#1}}}
\newcommand{\lEq}[1]{{\label{eq:#1}}}

\def\gtaprx {\lower .1ex\hbox{\rlap{\raise .6ex\hbox{\hskip .3ex
	{\ifmmode{\scriptscriptstyle >}\else
		{$\scriptscriptstyle >$}\fi}}}
	\kern -.4ex{\ifmmode{\scriptscriptstyle \sim}\else
		{$\scriptscriptstyle\sim$}\fi}}}
\def\ltaprx {\lower .1ex\hbox{\rlap{\raise .6ex\hbox{\hskip .3ex
	{\ifmmode{\scriptscriptstyle <}\else
		{$\scriptscriptstyle <$}\fi}}}
	\kern -.4ex{\ifmmode{\scriptscriptstyle \sim}\else
		{$\scriptscriptstyle\sim$}\fi}}}
\newcommand{\FIGFF}[2]{{\ref{fig:#2}{#1}}}

\newcommand{\FIG}[2]{{Fig.~\FIGFF{#1}{#2}}}
\newcommand{\Fig}[1]{{\FIG{}{#1}}}

\newcommand{\Sectff}[1]{{\ref{sec:#1}}}
\newcommand{\Sect}[1]{{\S~\Sectff{#1}}}

\newcommand{\Eqref}[1]{{\ref{eq:#1}}}

\newcommand{\eqff}[1]{{\Eqref{#1}}}

\newcommand{\eq}[1]{{equation~\eqff{#1}}}

\newcommand{\Msun}{\mbox{$M_\odot$}}

\begin{document}


\title{Bright Supernovae from Magnetar Birth}

\author{S. E. Woosley\altaffilmark{1}}

\altaffiltext{1}{Department of Astronomy and Astrophysics, University
  of California, Santa Cruz, CA 95064; woosley@ucolick.org}

\begin{abstract} 
Following an initial explosion that might be launched either by
magnetic interactions or neutrinos, a rotating magnetar radiating
according to the classic dipole formula could power a very luminous
supernova. While some $^{56}$Ni might be produced in the initial
explosion, the peak of the light curve in a Type I supernova would not
be directly related to its mass. In fact, the peak luminosity would be
most sensitive to the dipole field strength of the magnetar. The tail
of the light curve could resemble radioactive decay for some time but,
assuming complete trapping of the pulsar emission, would eventually be
brighter. Depending on the initial explosion energy, both high and
moderate velocities could accompany a very luminous light
curve.

\end{abstract}


\section{INTRODUCTION}
\lSect{intro}

The role of rotation in powering the explosion of supernovae has long
been debated \citep[e.g.,][]{Hoy46,Leb70,Ost71,Aki03}.  Most recent
work has focused on the possibility that a rotating neutron star could,
by way of a magnetic interaction, be the energy source for {\sl
  exploding} a massive star. While that issue is far from resolved,
and recent headway has been made in exploding these same stars using
neutrino transport, it is certain that a large fraction of supernova
explosions produce rotating neutron stars and that those neutron stars
frequently have large magnetic fields. Magnetars are a class of
neutron stars with field strengths 10$^{14}$ to 10$^{15}$ Gauss and
more. They may constitute 10\% of the neutron star birthrate
\citep{Kou98}. It is likely that fast rotation in the collapsing iron
core is responsible for creating the large magnetic field
\citep{Dun92}, so it is reasonable to expect that the birth of rapidly
rotating, highly magnetic neutron stars is commonplace.

It is also generally assumed that these rapidly rotating neutron stars
are magnetically braked by dipole emission early on, accounting for
the slow periods observed in anomalous x-ray pulsars and soft
gamma-ray repeaters \citep{Dun92,Kou98}. Since the initial rotational
energy of such stars at birth must have been large and 1000 years
later is small, where did the difference go? Might it have been
emitted in some observable form?

As a rough approximation, assume that the neutron star radiates its
rotational energy away at a rate given by the traditional dipole
formula for pulsars.  For a typical moment of inertia of 10$^{45}$ g
cm$^2$, the rotational energy of a neutron star with period, P$_{\rm
  ms}$, in milliseconds is
\begin{equation}
\begin{split}
E &= \frac{1}{2} I \omega^2 \\
&\approx 2 \times 10^{52} {\rm P}_{\rm ms}^{-2} \ {\rm erg}.
\end{split}
\lEq{pulsener}
\end{equation}
The approximate energy loss for dipole radiation is given by the Larmor
formula \citep[e.g.,][]{Lan80},
\begin{equation}
\begin{split}
\frac{d E}{d t} &= \frac{2}{3 c^3} \left(B R^3 \ {\rm Sin} \, \alpha \right)^2
\left(\frac{2 \pi}{{\rm P}}\right)^4 \\
&\approx 10^{49} B_{15}^2 {\rm P}_{\rm ms}^{-4} \ {\rm erg \ s^{-1}}.
\end{split}
\lEq{dipole}
\end{equation}
Here $B_{15}$ is the surface dipole field in 10$^{15}$ Gauss, $R
\approx 10^6$ cm is the neutron star radius, and $\alpha$ is the
inclination angle between the magnetic and rotational axes, taken
arbitrarily to be 30 degrees.

At face value, these simple formulae suggest large luminosities for
magnetars during the first few weeks of their lives. For example, the
magnetar in SGR 1627-41 is estimated to have a current spin-down
luminosity of $\sim 4 \times 10^{34}$ erg s$^{-1}$ at an age of 2.2 ky
implying a magnetic field $B \, {\rm Sin} \, \alpha \ \sim 2 \times
10^{14}$ Gauss \citep{Esp09}. Extrapolated back to when the magnetar
was 10 days old, \eq{dipole} implies a luminosity of $2 \times 10^{44}$
erg s$^{-1}$.  Provided the initial rotation rate was rapid compared
with its value at 10 days, this result is independent of the initial
rotation rate.

Here we explore the observational characteristics of Type Ib/c
supernovae in which pulsars with magnetar-like characteristics have
been embedded. The brightest supernovae actually come from neutron
stars with fields that, for a magnetar, are relatively modest, $\sim
10^{14}$ Gauss. Larger fields imply that most of the magnetar's
rotational energy is dissipated early on, adding to the kinetic energy
and mixing, but not appreciably affecting the luminosity.  These
luminous magnetar-powered supernovae might easily be confused with
other forms of ``hypernovae'' where the luminosity has a radioactive
origin. Alternatively, the lack of such emission constrains the
existence of any rapidly rotating magnetar. This could be an important
constraint in the context of gamma-ray bursts where the the
possibility of a magnetar power source is currently debated.

\section{A Model Explosion}

The presupernova model adopted for our first study is a 4.37 \Msun
\ Wolf-Rayet star derived from the evolution of single 35 \Msun \ main
sequence star \citep{Woo07}. Magnetars may originate from a population
this massive or larger \citep{Mun06}, and the relatively small helium
core mass will demonstrate the possibility of a very luminous
supernova from a light progenitor. A heavier model is explored later.
This star was exploded with a piston located at 1.89 \Msun, the base
of the former oxygen burning shell where the entropy per baryon was
$S/N_A k$ = 4.0. The piston was moved with a speed such as to impart a
final kinetic energy (without any pulsar addition) of $1.2 \times
10^{51}$ erg and produce 0.05 \Msun \ of $^{56}$Ni (\Fig{comp}). The
specific nature of this initial explosion is left unspecified. It
could have been the result of successful neutrino-mediated transport
or a consequence of early energy deposition related to the formation
of the magnetar itself.

\begin{figure}
\includegraphics[width=0.47\textwidth]{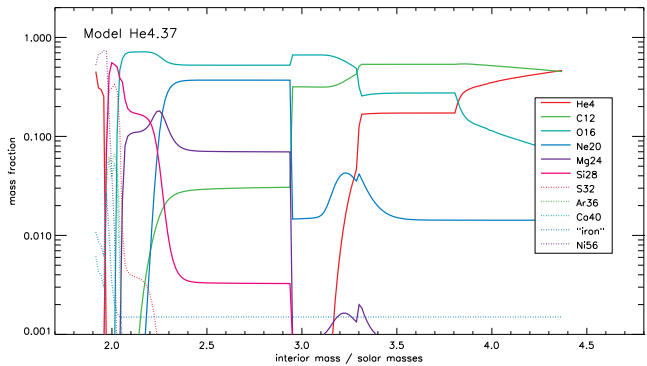}
\caption{Final composition of the explosion whose light curve was
  shown in \Fig{lite}. The kinetic energy of the explosion is $1.2
  \times 10^{51}$ erg without the pulsar, $1.6 \times 10^{51}$ erg
  with it.  \lFig{comp}}
\end{figure}

Ten seconds after the explosion, when all important nucleosynthesis
had ceased and the shock wave was still about 0.5 \Msun \ beneath the
stellar surface, energy deposition from an assumed embedded magnetar
was turned on. A magnetic field of 10$^{14}$ Gauss and an initial
rotation period of 4.5 ms were assumed, corresponding to a
rotational kinetic energy of 10$^{51}$ erg.  Energy from the magnetar
was deposited in the inner 10 zones of the model at a rate given by
\eq{dipole} and complete trapping was always assumed. While the form
of the energy deposition was unspecified, it could be accelerated
particles or radiation of any wavelength to which the supernova is
initially opaque. The initial luminosity of the assumed pulsar was
10$^{45}$ erg s$^{-1}$ (\eq{dipole}) which declined to 10$^{42}$ erg
s$^{-1}$ after $5.9 \times 10^7$ s.  After $2 \times 10^{11}$ s the
rotation period was 1.0 s.

The resulting light curve is shown in \Fig{lite}.  At early times the
effect of the magnetar was negligible. Later, some of the energy went
into accelerating the ejecta to higher speeds and creating a
``bubble'' inside the supernova. Later the energy was mostly radiated
away and, at late times, the supernova luminosity tracked \eq{dipole}.
The light curve peaked at $7.3 \times 10^{43}$ erg s$^{-1}$ (about 5
typical Type Ia supernovae) at 37 days. At that time, the pulsar was
depositing $7.5 \times 10^{43}$ erg s$^{-1}$. The rise time from
10$^{42}$ erg s$^{-1}$ was 24 days.

\begin{figure}
\includegraphics[angle=90,width=0.47\textwidth]{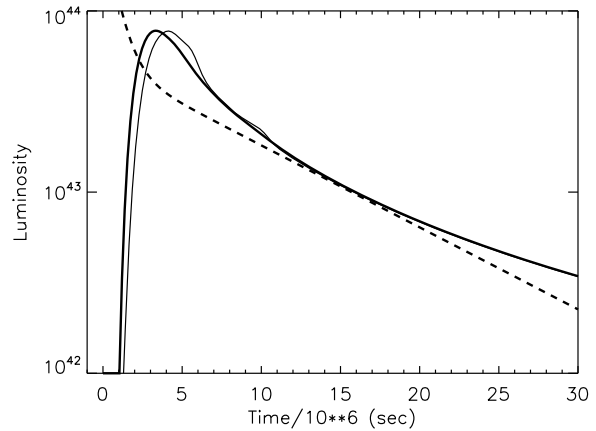}
\caption{Light curve of a 4.37 \Msun \ Wolf-Rayet star in which a
  magnetar (10$^{14}$ Gauss; $E_o$ = 10$^{51}$ erg) emitting dipole
  radiation is embedded. Complete trapping of the magnetar emission is
  assumed at all times. Shown for comparison is the energy released by
  the decay of 4 \Msun \ of $^{56}$Ni. If the light curve at peak had
  been powered by radioactivity it would have required 7 \Msun. The
  second solid line, slightly displaced to later times, shows the
  light curve of a heavier star (7.29 \Msun) exploded with the same
  energy and containing the same magnetar (see
  \Sect{heavy}). \lFig{lite}}
\end{figure}

Also of some interest is now the magnetar emission modifies the
dynamics of the explosion. Without the pulsar, the explosion energy
was $1.2 \times 10^{51}$ erg; with it the explosion energy is
increased to $1.6 \times 10^{51}$ erg. That is about 40\% of the
magnetar energy went into kinetic energy of expansion, while the
remainder went into radiation. Energy was not deposited uniformly in
the ejecta though. Because the pulsar was a central source of heating,
its energy went into inflating a bubble of lower density material
inside the supernova. The boundary of this bubble is a dense shell in
which resides about a solar mass of the ejecta (\Fig{vel}). Other
calculations, not illustrated here, show that if one assumes a larger
magnetar rotational energy or stronger magnetic field, the velocity
and mass of this thin shell are increased.

In three dimensions, this shell is probably unstable and, rather than
being ejected as a thin spherical shell, the ejecta are probably
mixed. This will affect the light curve, perhaps making it rise at an
earlier time. Multi-dimensional studies of the coupled radiation
transport and hydrodynamics are needed and are feasible, but are
postponed for now. This mixing will also have important implications
for the spectrum, especially at late times.

\begin{figure}
\includegraphics[angle=90,width=0.47\textwidth]{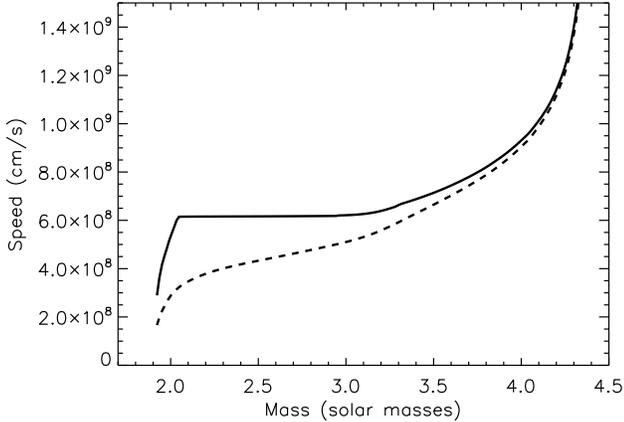}
\caption{Final velocity for two explosions in which the pulsar energy
  deposition was included (solid line) or was not (dashed line). In
  one dimension, the energy deposited in the inner ten zones blows a
  bubble resulting in a high density shell moving at nearly constant
  speed. In three dimensions, this shell is probably unstable and
  mixing would occur.  \lFig{vel}}
\end{figure}

\section{A Heavier Progenitor and a Stronger Field}
\lSect{heavy}

The same magnetar in supernovae with different masses and kinetic
energies will give supernovae with different properties. Consider
first the effect of the same magnetar as in the previous section
($E_{\rm rot} = 10^{51}$ erg, $B_{15} = 0.1$) embedded in a Type Ib
supernova with mass 7.29 \Msun. The presupernova star here was the
remnant of a 60 \Msun \ main sequence star of solar metallicity, again
taken from the study of \citet{Woo07}. The star was exploded, as
before, with a piston at the base of the oxygen shell (1.59 \Msun) and
given a final kinetic energy of $1.2 \times 10^{51}$ erg. Explosive
nucleosynthesis produced 0.11 \Msun \ of $^{56}$Ni though the effect
of $^{56}$Ni decay was {\sl not} included in the plots shown.

\Fig{lite} already showed the resulting light curve. It is very nearly
the same as for the lower mass model. With its larger mass and similar
kinetic energy, the supernova expands slower, but the emerging light
is only slightly delayed and on the tail, the light curves are
identical.

\Fig{lite2} shows what happens if the magnetar is assumed to have a
stronger magnetic field, $B_{15}$ = 0.5 and 0.7. More of the
rotational energy is deposited early on and contributes to the
expansion rate. Consequently the light curve at peak is fainter and it
declines more rapidly. Smaller initial rotation rates ($2 \times
10^{50}$ and $5 \times 10^{50}$ erg corresponding to periods of 10 ms
and 6.3 ms) were employed, but the answer is not very sensitive to
that. The extra energy would just accelerate the ejecta. Since the
resulting thin spherical shells are probably not physical, they are
not considered here.

\begin{figure}
\includegraphics[angle=90,width=0.47\textwidth]{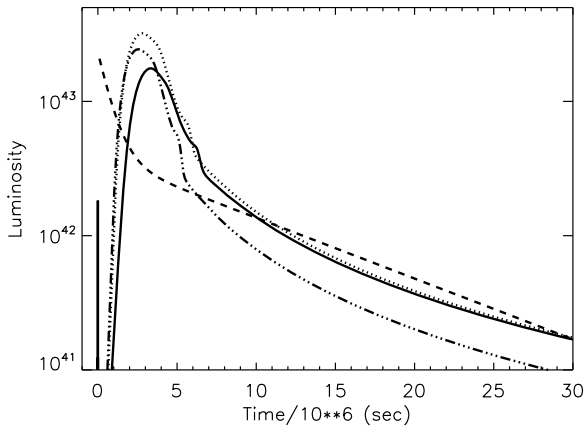}
\caption{Light curves for three models based upon the 1.2 $\times
  10^{51}$ erg explosion of a 7.29 \Msun \ Wolf-Rayet star. The
  embedded magnetar had a field strength of $5 \times 10^{14}$ Gauss
  and initial rotational energies of $2 \times 10^{50}$ (solid line)
  and $5 \times 10^{50}$ erg (dotted line) or $7 \times 10^{14}$
  Gauss and $5 \times 10^{50}$ erg (dot-dashed line). The dashed line
  shows the energy that would result from the decay of 0.3 \Msun \ of
  $^{56}$Ni.  \lFig{lite2}}
\end{figure}

The light curves in \Fig{lite2} resemble, superficially at least, that
of SN 1998bw which accompanied GRB 980425 \citep{Pat01}. The long
tails that in fact go like $t^{-2}$, parallel for some time that
expected from the decay of $^{56}$Co and if the light curve for the
first year was all one had to go by, one might easily confuse a
pulsar-powered model for one powered by radioactivity. Since the
explosion energy and mechanism is left unspecified here, one could
invoke a much higher kinetic energy (accompanying the birth of the
magnetar?) and obtain spectra like that seen in SN 1998bw. The light
curve at peak would then be powered by a combination of radioactivity
and pulsar emission.

However, the luminosity of SN 1998bw at times later than one year
\citep{Sol02} argues against such a hybrid model. By 1000 days after
the explosion, the supernova had declined to $\sim10^{38}$ erg
s$^{-1}$, whereas a pulsar with field strength $5 \times 10^{14}$
Gauss would still be radiating $2 \times 10^{40}$ erg s$^{-1}$. In
fact, a strict application of \eq{dipole} would imply that the pulsar
must have been born with a field strength in excess of $8 \times
10^{15}$ Gauss in order to satisfy the observational limit at late
times, but then the magnetar would contribute all of its rotational
energy to the explosion and none to the light curve.

Of course \eq{dipole} is an approximation whose validity can be
questioned, especially at early times, and the magnetic field strength
and orientation need not be constant with time. Given the freedom to
pick a field decay rate, the entire light curve could be fit, but this
seems somewhat contrived. Fitting the late time light curve with a
combination of $^{56}$Co and $^{57}$Ni decay seems, for now, more
natural \citep{Sol02}.

\section{Conclusions}

Pulsars with magnetar-like magnetic fields can contribute appreciably
to the light curves of Type Ib and Ic supernovae. In fact, it would be
surprising if they never did. For moderate field strengths, the light
curve is very luminous and long lasting and might be confused with
those of pair-instability supernovae or circumstellar
interaction. Indeed, \citet{Mae07} have suggested a pulsar as the
possible source powering the second (principal) maximum of the light
curve of SN 2005bf.

If magnetars are the central engine that powers the long-soft class of
gamma-ray bursts \citep[e.g.,][]{Woo06}, then it is reasonable to
expect that they may contribute to the light curves of the supernovae
that accompany them. On the one hand, this might facilitate the
magnetar paradigm because the production of the necessary large amount
of $^{56}$Ni has proven problematic \citep{Buc09}. Having an alternate
explanation that involves magnetar energy input would solve this
problem. On the other hand, if a magnetar contribution to the light
curve can be ruled out based on the spectrum and late time light
curve, one must wonder how the magnetar is so effectively
concealed. Is the rotational energy extracted with such high
efficiency in the first few seconds that the magnetar forever
afterwards rotates slowly, or do rotating magnetars not emit according
to the popular dipole formula during the first year?

Perhaps they do not. The absence of a pulsar contribution to the
luminosity of typical supernovae is easy to understand. A pulsar born
with a 14 ms period (10$^{50}$ erg), would only have a pulsar
luminosity of $6 \times 10^{39}$ erg s during the time when it is
bright. This is small compared with the 10$^{42}$ - 10$^{43}$ erg
s$^{-1}$ resulting from shock energy released by recombination (Type
II supernova) or radioactivity (Type I supernova). But the extremely
faint optical luminosity of SN 1987A, $< 8 \times 10^{33}$ erg
s$^{-1}$ seventeen years after its birth \citep{Gra05} is very
difficult to reconcile with any model with a young active
pulsar. While dust extinction could be considerable, the lack of a
point source in SN 1987A remains a mystery.

\acknowledgements

The author gratefully acknowledges helpful conversations with Lars
Bildsten, Dan Kasen and Paulo Mazzali. This research has been supported
by the NASA Theory Program (NNG05GG08G) and the DOE SciDAC Program
(DE-FC02-06ER41438).

\clearpage

\end{document}